# An Investigation of the Variation in Seasonal Rainfall Patterns Over the Years

Priti Kaushik, Randhir Singh Baghel

Faculty of Science, Poornima University, Jaipur- 303905, Rajasthan, India

E-mail: priti.kaushik@poornima.edu.in, randhir.baghel@poornima.edu.in,

**Abstract**

In this work, long-term spatiotemporal changes in rainfall are analyzed and evaluated using whole-year data from Rajasthan, India, at the meteorological divisional level. In order to determine how the rainfall pattern has changed over the past 10 years, I examined the data from each of the thirteen tehsils in the Jaipur district. For the years 2012 through 2021, daily rainfall information is available from the Indian Meteorological Department (IMD) in Jaipur. We primarily compare data broken down by tehsil in the Jaipur district of Rajasthan, India.

## INTRODUCTION

The biggest capital of Rajasthan is Jaipur, sometimes referred to as the "Pink City." Maharaja Sawai Jai Singh II, the ruler of Amber, began construction on it on November 18, 1727. District Jaipur is 390 meters above sea level. Historic forts including Amber Fort, Jaigarh Fort, and Hawa Mahal, among others, are well recognized for being in Jaipur City. Jaipur district has an area of 11,117 sq km, with a population density of 470 people per sq km, while Jaipur city is a total of 467 sq. km [1-6]. The district is around 180 kilometres long from east to west and 110 km wide from north to south. There are thirteen tehsils: Amber, Viratnagar, Bassi, Chaksu, Chomu, Phulera, Phagi, Dudu, Kotputli, Jaipur, Jamuwa Ramgarh, Sanganer, and Shahpura. Its borders are formed by the districts of Sikar and Mahendrakar to the north, Tonk district to the south, Alwar, Sawai Madhapur, and Dauda district to the east, and Nagaur district to the west. According to the 2011 census, 3.9 million people are living in Jaipur overall.

The Jaipur district is located on the eastern edge of the semi-arid Thar desert. A far-off location from the Arabian Sea and the Bay of Bengal creates a continental climate. In the aftermath of western disturbances, humidity, cloudiness, and rainfall activities increase throughout the monsoon season from July to September and sporadically during the off-season [7–15].

The year is roughly split into four seasons: the winter season, which lasts from December to February; the summer or hot weather season, which lasts from March to May; the monsoon season, which lasts from the end of June to mid-September; and the transit period, which lasts from October to November.

Winters in Jaipur are quite cold, while summers are extremely hot. The hottest temperatures in May normally fall between 40 °C and 47 °C. When daytime temperatures reach 4 to 6 °C above average for a few days throughout the season, a heat wave develops [16–17].

Winter lows typically range from 4 to 9 °C and are below 0 degrees. or when a cold, northerly wind comes out of the Himalayan region. Occasionally in the morning hours after the passage of western disturbances, mist and fog also appear [18-24].

Ecology enhances the planet and is essential to the well and prosperity of people [25-27]. It offers a fresh understanding of the connection between humans and the environment, which is essential for food production, preserving clean air and water, and preserving biodiversity in the face of climate change [28-30].

**The area of different Tehsils is given below**

| Sr. no | Tehsils | Area (sq. km.) |
|---|---|---|
| 1 | Amber | 891.22 |
| 2 | Bassi | 654.69 |
| 3 | Chaksu | 811.77 |
| 4 | Chomu | 683.61 |
| 5 | Dudu | 1338.56 |
| 6 | Jaipur | 467 |
| 7 | Jamuwa Ramgarh | 1033.7 |
| 8 | Kotputli | 814.34 |
| 9 | Phagi | 1114.34 |
| 10 | Phulera | 1470.38 |
| 11 | Sanganer | 701.75 |
| 12 | Shahpura | 530.96 |

# JAIPUR

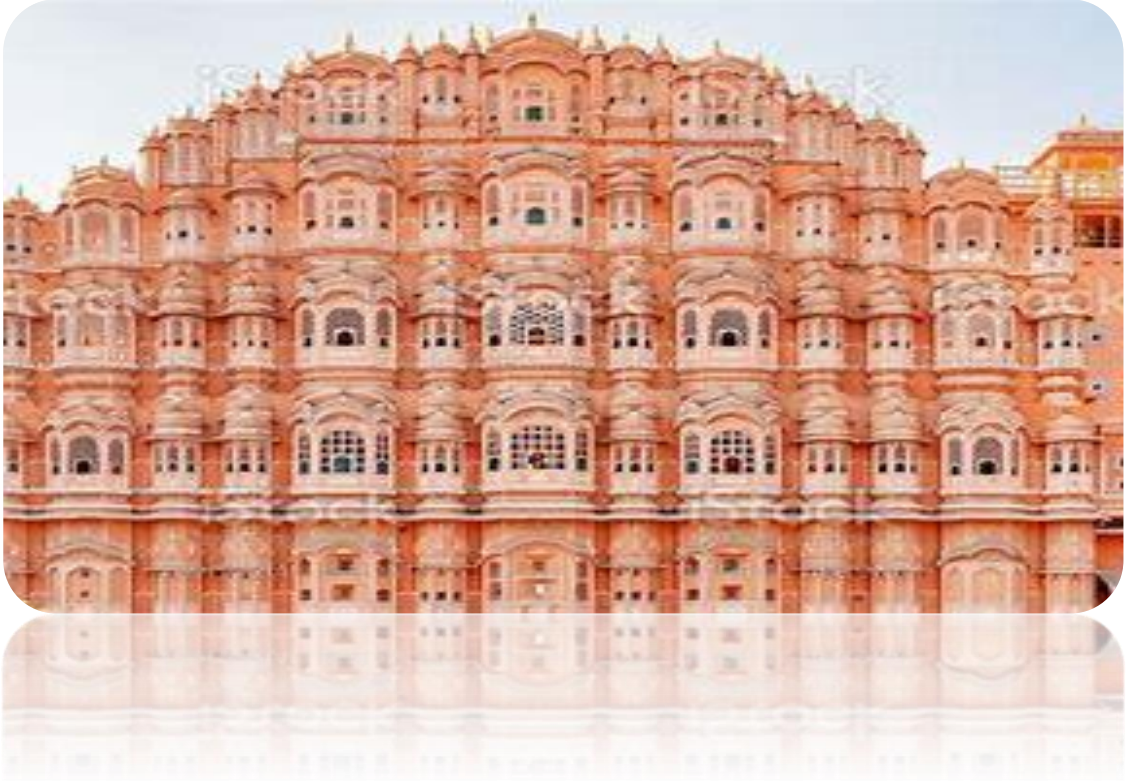

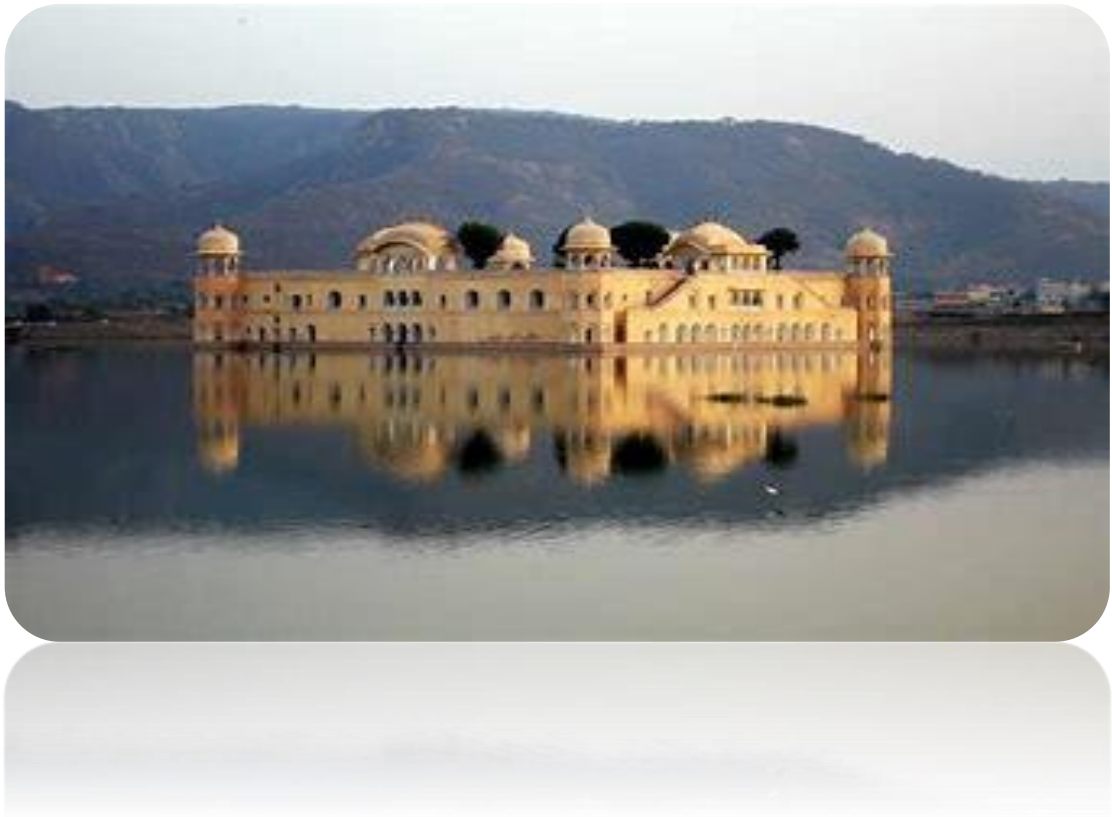

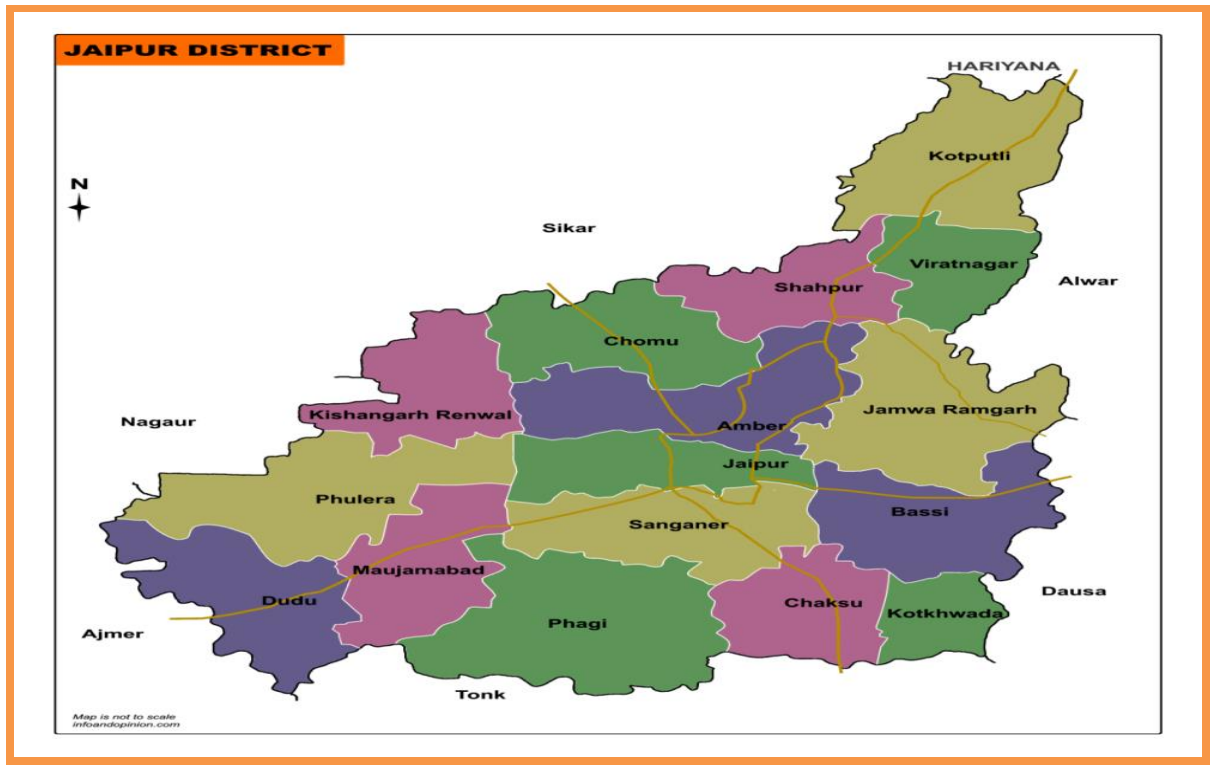

**DISTRICT MAP OF JAIPUR**

## METHODOLOGY AND DATA USED

The statistics from each of the thirteen thesils in the Jaipur district were studied to see how the rainfall pattern has evolved over the previous 10 years. Daily rainfall data for the years 2012 through 2021 have been made available by the Indian Meteorological Department (IMD) in Jaipur. After obtaining the information, make sure the date and year are appropriately placed and check to see if any values are missing. The weekly, monthly, and seasonal rainfall of Jaipur tahsil were calculated using this structured data by using Python and replacing any null spaces with 0 to guarantee that the computations would be error-free.

For data analysis, both Python and Microsoft Excel have been employed. to organize the rainfall data for each tehsil to determine the maximum rainfall.

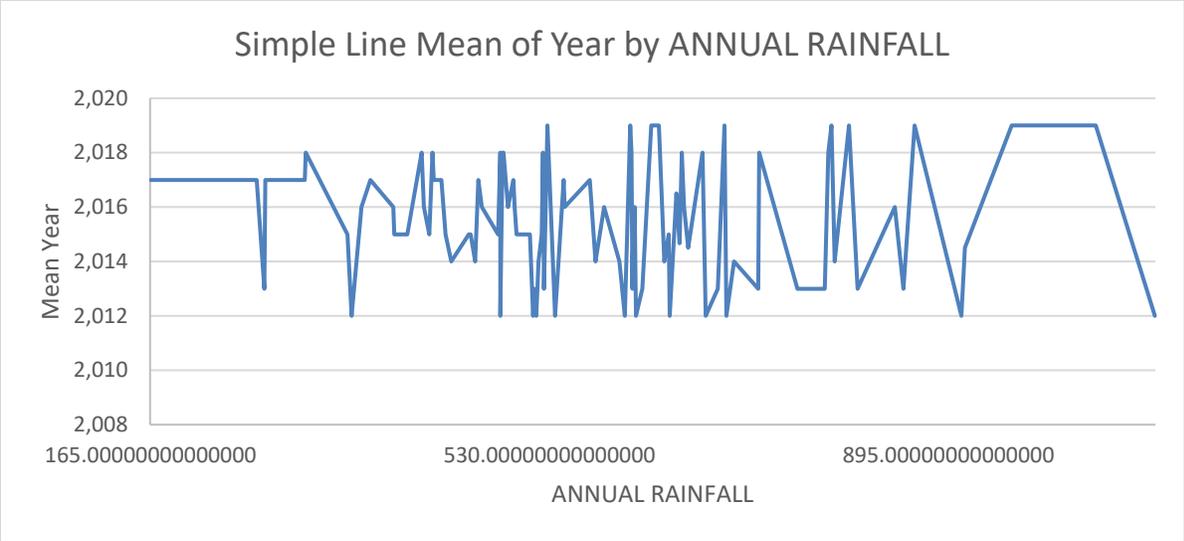

## Time Series Modeler

| Model Description | | | |
|---|---|---|---|
| | | | Model Type |
| Model ID | Jan | Model_1 | ARIMA(0,0,16) |
| | Feb | Model_2 | ARIMA(0,0,8) |
| | March | Model_3 | ARIMA(0,0,16) |
| | April | Model_4 | ARIMA(2,0,4) |
| | May | Model_5 | ARIMA(0,0,0) |
| | Jun | Model_6 | ARIMA(0,0,16) |
| | July | Model_7 | ARIMA(0,0,8) |
| | Aug | Model_8 | ARIMA(0,0,5) |
| | Sep | Model_9 | ARIMA(0,0,8) |
| | Oct | Model_10 | ARIMA(0,0,4) |
| | Nov | Model_11 | Simple |
| | Dec | Model_12 | ARIMA(0,0,8) |

| Model Statistics | | | | | | |
|---|---|---|---|---|---|---|
| | | Model Fit Statistics | Ljung-Box Q(18) | | | |
| Model | Number of Predictors | Stationary R-squared | Statistics | DF | Sig. | Number of Outliers |
| Jan-Model_1 | 0 | .281 | 12.865 | 16 | .683 | 0 |
| Feb-Model_2 | 0 | .071 | 24.884 | 17 | .097 | 0 |
| March-Model_3 | 0 | .532 | 27.495 | 16 | .036 | 0 |
| April-Model_4 | 0 | .263 | 34.036 | 15 | .003 | 0 |
| May-Model_5 | 0 | 7.772E-16 | 17.071 | 18 | .518 | 0 |
| Jun-Model_6 | 0 | .241 | 14.372 | 16 | .571 | 0 |
| July-Model_7 | 0 | .192 | 37.548 | 17 | .003 | 0 |
| Aug-Model_8 | 0 | .104 | 66.075 | 17 | .000 | 0 |
| Sep-Model_9 | 0 | .147 | 54.490 | 17 | .000 | 0 |
| Oct-Model_10 | 0 | .069 | 11.644 | 17 | .821 | 0 |
| Nov-Model_11 | 0 | .389 | 7.871 | 17 | .969 | 0 |
| Dec-Model_12 | 0 | .202 | 9.536 | 16 | .890 | 0 |

| Exponential Smoothing Model Parameters | | | | | | |
|---|---|---|---|---|---|---|
| Model | | | Estimate | SE | t | Sig. |
| Nov-Model_11 | No Transformation | Alpha (Level) | .038 | .017 | 2.226 | .028 |

## ARIMA Model Parameters

| | Model Fit | | | | | | | | | | | Sig. |
|---|---|---|---|---|---|---|---|---|---|---|---|---|
| | | | | | | | Percentile | | | | | .000 |
| Fit Statistic | Mean | SE | Minimum | Maximum | 5 | 10 | 25 | 50 | 75 | 90 | 95 | .000 |
| Stationary R-squared | .208 | .149 | 7.772E-16 | .532 | 7.772E-16 | .021 | .079 | .197 | .276 | .489 | .532 | |
| R-squared | .172 | .150 | -.020 | .532 | -.020 | -.014 | .070 | .160 | .257 | .457 | .532 | .000 |
| RMSE | 32.420 | 39.544 | 2.792 | 122.411 | 2.792 | 2.993 | 9.078 | 13.377 | 44.740 | 115.635 | 122.411 | .001 |
| MAPE | 116.351 | 63.805 | 48.132 | 247.156 | 48.132 | 56.267 | 78.807 | 86.597 | 134.294 | 244.664 | 247.156 | |
| MaxAPE | 1319.210 | 1460.408 | 98.597 | 4531.352 | 98.597 | 136.218 | 436.254 | 722.741 | 1773.179 | 4367.354 | 4531.352 | .034 |
| MAE | 24.102 | 30.481 | 1.321 | 96.800 | 1.321 | 1.452 | 6.906 | 8.504 | 35.509 | 89.012 | 96.800 | .002 |
| MaxAE | 121.162 | 143.994 | 13.123 | 437.917 | 13.123 | 15.918 | 31.483 | 54.624 | 146.976 | 423.942 | 437.917 | |
| Normalized BIC | 5.765 | 2.406 | 2.188 | 9.704 | 2.188 | 2.290 | 4.535 | 5.260 | 7.702 | 9.582 | 9.704 | .000 |

| | | | | | | | | | Sig. |
|---|---|---|---|---|---|---|---|---|---|
| | | | | Lag 16 | | | -.250 | .106 | -2.357 | .020 |
| April-Model_4 | April | No Transformation | Constant | | | 8.105 | 1.013 | 8.002 | .000 |
| | | | AR | Lag 2 | | -.923 | .056 | -16.594 | .000 |
| | | | MA | Lag 2 | | -.956 | .112 | -8.556 | .000 |
| | | | | Lag 4 | | -.242 | .104 | -2.325 | .022 |
| May-Model_5 | May | No Transformation | Constant | | | 8.028 | .925 | 8.682 | .000 |
| Jun-Model_6 | Jun | No Transformation | Constant | | | 51.421 | 7.281 | 7.063 | .000 |
| | | | MA | Lag 8 | | -.352 | .102 | -3.435 | .001 |
| | | | | Lag 16 | | -.480 | .115 | -4.156 | .000 |
| July-Model_7 | July | Natural Logarithm | Constant | | | 5.152 | .067 | 76.652 | .000 |
| | | | MA | Lag 8 | | -.438 | .100 | -4.365 | .000 |
| Aug-Model_8 | Aug | No Transformation | Constant | | | 208.276 | 8.052 | 25.868 | .000 |
| | | | MA | Lag 5 | | .344 | .095 | 3.634 | .000 |
| Sep-Model_9 | Sep | No Transformation | Constant | | | 61.966 | 5.814 | 10.658 | .000 |
| | | | MA | Lag 8 | | -.344 | .098 | -3.512 | .001 |
| Oct-Model_10 | Oct | No Transformation | Constant | | | 8.871 | .860 | 10.315 | .000 |
| | | | MA | Lag 4 | | .258 | .097 | 2.659 | .009 |
| Dec-Model_12 | Dec | No Transformation | Constant | | | 1.281 | .457 | 2.805 | .006 |
| | | | MA | Lag 1 | | -.321 | .089 | -3.599 | .000 |
| | | | | Lag 8 | | -.381 | .099 | -3.853 | .000 |

**Discussion & Conclusion:**

Based on the above data, there are four rainy seasons throughout the year (winter monsoon season, pre-monsoon season, monsoon season, and post-monsoon season). The winter monsoon season lasts two months (January and February), the pre-monsoon season lasts three months (March, April, and May), the main monsoon season lasts four months (June, July, August, and September), and the post-monsoon season lasts three months (October, November, and December). The monsoon season (June, July, August, and September) receives the most rainfall of any of the four rainy seasons. When total rainfall for each annual is compared, it can be seen that July and August always receive more rainfall than any other month. The amount of rainfall typically increases starting in week 25th and peaks in week 32nd or week 33rd. Once a year, rainfall crosses the barrier of 114.5 mm, while rainfall beyond the barrier of 204.5 mm is extremely rare. In many cases, the year 2017 has received the lowest rainfall during the long period of 10 years. During the period 2012-2021, the years 2013, 2019, and 2020 experienced the most rainfall. In contrast, Kotputli experiences more rainfall in July than in August. We have used SPSS to analyze the rainfall forecast.

We may deduce from this study's comparison of every tehsil in the Jaipur district that Phagi experienced 1087 mm of yearly rainfall in 2021, which is 3 mm higher than the city of Jaipur as a whole did in 2012. The annual rainfall in Jaipur in 2012 was around 1084 mm. Jaipur received the highest one-day rainfall total throughout the analysis period, which is far more than any other tehsil at almost 300 mm. Table 19.2 shows that rainfall more than 204.5 mm or exceptionally heavy rainfall is exceedingly unusual, comparable to a super moon since it only occurs four times in 10 years.